\documentclass[12pt]{article}
\usepackage{latexsym}
\newcommand{\beq}{\begin{equation}}
\newcommand{\eeq}[1]{\label{#1}\end{equation}}
\newcommand{\bea}{\begin{eqnarray}}
\newcommand{\eea}[1]{\label{#1}\end{eqnarray}}

\begin{document}
\hfill NYU-TH/00/11/02 hep-th/0011152

\vspace{20pt}

\begin{center}
{\textbf{NO VAN DAM-VELTMAN-ZAKHAROV DISCONTINUITY IN ADS SPACE}}
\end{center}

\vspace{6pt}

\begin{center}
\textsl{M. Porrati} \vspace{20pt}

\textit{Department of Physics, NYU, 4 Washington Pl, New York NY 10003}

\end{center}

\vspace{12pt}

\begin{center}
\textbf{Abstract }
\end{center}

\vspace{4pt} {\small \noindent
We prove that the van Dam-Veltman-Zakharov discontinuity arising in the 
massless limit of massive gravity theories is peculiar to Minkowski space and
it is not present in Anti De Sitter space, where the massless limit is smooth.
More generally, the massless limit is smooth whenever the square of the
graviton mass vanishes faster than the cosmological constant.

\vfill\eject 
\noindent
In flat space, the massless limit of a massive spin-2 field coupled to the 
covariantly conserved stress-energy tensor 
yields a massless spin-2, a massless vector, and a
massless spin-0 that couples with gravitational strength to the trace of the
stress-energy tensor. This is the origin of the famous van
Dam-Veltman-Zakharov
(vDVZ) discontinuity~\cite{vvdz} (see also~\cite{bd} and references therein): 
because of the extra scalar massive gravity
predicts a value for the gravitational bending of light by a massive source 
that is 3/4 of the Einstein prediction. This result extends to any ghost-free 
theory of massive spin-2 coupled to the stress-energy tensor~\cite{dgp}
(see also~\cite{grs,c,prz,kro} for related discussions).

Recently a unexpected result has been obtained by Karch and Randall~\cite{kr},
who studied the graviton propagator in a warped compactification similar 
to~\cite{rs}, but where the four dimensional metric is Anti de Sitter
instead of Minkowski.  Karch and Randall find that in their 
compactification the graviton is massive, with mass $O(\Lambda^2)$, yet the
limit $\Lambda \rightarrow 0$ is smooth and gives the usual RS propagator
of flat space~\cite{rs,gkr}. In this paper $\Lambda$ is the ``cosmologist's'' 
one, related to the vacuum energy $V$ by $\Lambda=8\pi G_{Newton} V$.
The smoothness of the $\Lambda\rightarrow 0$ limit seems in 
contradiction with the existence of a vDVZ discontinuity. 

In this paper, we examine the propagation of massive spin-2 in AdS space, and
compute the one-particle amplitude between covariantly conserved sources.
We find that, contrary to flat-space expectation, when $\Lambda<0$ there is no

discontinuity in the massless limit. More precisely, the one-particle 
exchange amplitude converges to the massless one when 
$M^2/\Lambda\rightarrow 0$.  This is consistent with the the findings of 
Karch and Randall. 

More than ten years ago, Higuchi~\cite{h} showed that there is no vDVZ 
discontinuity in {\em de Sitter} space, i.e. at $\Lambda >0$. Since in that 
case there is no unitary spin-2  representation in the mass range 
$0<M^2<2\Lambda/3$~\cite{h} one could have interpreted that result as due to 
the presence of ghosts. In Anti de Sitter space, instead, spin-2 unitary 
representations of the AdS group exist for all $M^2\geq 0$. 
Higuchi's proof makes
use of the explicit form of the propagator in de Sitter space, so it is not
immediately applicable to AdS. The method we use, instead, works for either
signs of the cosmological constant.   

Proof of our claim is straightforward. We start with the unique
ghost-free action for a free, massive spin-2 field propagating on an Einstein 
space, the Pauli-Fierz action~\cite{pf}
\beq
S=S_L[h_{\mu\nu}] + \int d^4x \sqrt{-g}\left[ {M^2\over 64\pi G_M}
(h_{\mu\nu}^2 -h^2) +{1\over 2}h_{\mu\nu}T^{\mu\nu}\right].
\eeq{1}
Here $S_L[h_{\mu\nu}]$ is the Einstein action with cosmological constant,
\beq
S_E[\hat{g}_{\mu\nu}]={1\over 16 \pi G_M} \int d^4x 
\sqrt{-\hat{g}}[R(\hat{g}) - \Lambda], \qquad \hat{g}_{\mu\nu} = 
g_{\mu\nu} + h_{\mu\nu}
\eeq{ein} 
linearized around the Einstein-space background $g_{\mu\nu}$. 
$G_M$ is the coupling 
constant of the spin-2 theory; by definition, $G_0=G_{Newton}$. 

All indices are raised, lowered and contracted with the 
background metric $g_{\mu\nu}$, and $T_{\mu\nu}$ is covariantly conserved
in the background.

By setting $8\pi G_M=2$, the equation of motion is
\beq
\Delta^{(2)}_Lh_{\mu\nu} + 2 \nabla_{(\mu}\nabla^\rho h_{\nu)\rho} -
\nabla_{(\mu}\nabla_{\nu)}h -2\Lambda h_{\mu\nu} +M^2h_{\mu\nu}
+{M^2\over 2}g_{\mu\nu} h =4T_{\mu\nu} -2 g_{\mu\nu}T.
\eeq{2}
In AdS, positive energy requires $M^2\geq 0$~\cite{ad}.

In Eq.~(\ref{2}), $\nabla_\mu$ is the covariant derivative with respect to the
background metric, and  $\Delta^{(2)}_L$ is the Lichnerowicz operator acting 
on spin-2 symmetric tensors~\cite{l}
\beq
\Delta^{(2)}_Lh_{\mu\nu}=-\nabla^2 h_{\mu\nu} -
2R_{\mu\rho\nu\sigma}h^{\rho\sigma} + 2R_{(\mu}^\rho h_{\nu)\rho}.
\eeq{3}
On the AdS background, $R_{\mu\rho\nu\sigma}=(\Lambda/3)(
g_{\mu\nu}g_{\rho\sigma} -g_{\mu\rho}g_{\nu\sigma})$, 
$R_{\mu\nu}=\Lambda g_{\mu\nu}$. On this background, the Lichnerowicz operator
obeys the following properties~\cite{l}:
\bea
\Delta^{(2)}_L \nabla_{(\mu} V_{\nu)}&=&\nabla_{(\mu}\Delta^{(1)}_L V_{\nu)},
\qquad \Delta^{(1)}_LV_\mu = (-\nabla^2 +\Lambda) V_\mu, 
\label{4} \\
\nabla^\mu \Delta^{(2)}_L h_{\mu\nu} &=& \Delta^{(1)}_L \nabla^\mu h_{\mu\nu},
\label{5} \\ 
\Delta^{(2)}_L g_{\mu\nu}\phi &=& g_{\mu\nu}\Delta^{(0)}_L\phi, \qquad 
\Delta^{(0)}_L\phi=-\nabla^2 \phi, 
\label{6}\\
\nabla^\mu \Delta^{(1)}_L V_\mu &=& \Delta^{(0)}_L \nabla^\mu V_\mu. 
\eea{7}

We will also need the following property, easily proven by using 
$[\nabla_\mu,\nabla_\nu]V_\rho=R_{\mu\nu\rho}^{\;\;\;\;\;\;\sigma} V_\sigma$
\beq
\nabla^{\mu}(\nabla_\mu V_\nu + \nabla_\nu V_\mu) = - \Delta^{(1)}_L V_\nu
+ 2\Lambda V_\nu +\nabla_\nu \nabla^\mu V_\mu.
\eeq{8}

First of all, let us check that Eq.~(\ref{2}) propagates only five degrees of 
freedom when $M^2>0$, and only two when $M^2=0$. This is a well known 
property, but the techniques introduced during the proof will be necessary 
to compute the one-particle amplitude.
\section*{$M^2 >0$}
Compute the double divergence of Eq.~(\ref{2}). Using Eqs.~(\ref{4}-\ref{8})
we find
\beq
(\nabla^2 + M^2) \nabla^\mu\nabla^\nu h_{\mu\nu} + (M^2/2 -\Lambda-\nabla^2)
\nabla^2 h =-2\nabla^2T.
\eeq{9}
Compute now the trace of Eq.~(\ref{2})
\beq
-2\nabla^2 h + 2\nabla^\mu \nabla^\nu h_{\mu\nu} -2\Lambda h + 3M^2 h =
-4T .
\eeq{10}
Apply $\nabla^2/2$ to Eq.~(\ref{10}) and subtract the result from 
Eq.~(\ref{9})
\beq
M^2(\nabla^\mu \nabla^\nu h_{\mu\nu} -\nabla^2 h)=0.
\eeq{11} 
Use Eq.~(\ref{11}) to eliminate $\nabla^\mu \nabla^\nu h_{\mu\nu}$ in
Eq.~(\ref{10}) and arrive to
\beq
(3M^2 -2\Lambda)h=-4T.
\eeq{12}
Setting $T=0$ we find $h=0$, whence $\nabla^\mu \nabla^\nu h_{\mu\nu}=0$. 

Compute now the divergence of Eq.~(\ref{2}). Using again
Eqs.~(\ref{4}-\ref{8})
we find
\beq
(\Delta^{(1)}_L +\nabla^2 -\Lambda) \nabla^\mu h_{\mu\nu} + \nabla_\nu
\nabla^\mu \nabla^\rho h_{\mu\rho} + \Lambda \nabla^\mu h_{\mu\nu}
- (\nabla^2 +\Lambda/2) \nabla_\nu h + M^2 \nabla^\mu h_{\mu\nu}
+{M^2\over 2} \nabla_\nu h =0.
\eeq{13}
Since $\nabla^\mu \nabla^\nu h_{\mu\nu}=h=0$, by using the definition of
$\Delta^{(1)}_L$ in Eq.~(\ref{4}), Eq.~(\ref{13}) reduces to
\beq
M^2 \nabla^\mu h_{\mu\nu} =0.
\eeq{14}
Eqs.~(\ref{12},\ref{14}) imply that on-shell $h_{\mu\nu}$ is transverse and 
traceless i.e. it propagates $10-4-1=5$ degrees of freedom.
\section*{$M^2=0$}
At $M^2=0$ Eq.~(\ref{2}) is invariant under the gauge transformation
$ h_{\mu\nu} \rightarrow h_{\mu\nu} + \nabla_{(\mu} V_{\nu)}$; using this
invariance to set
\beq
\nabla^\mu h_{\mu\nu}-{1\over 2}\nabla_\nu h=0,
\eeq{15}
setting $T_{\mu\nu}=0$, we find the equation
\beq
\Delta^{(2)}_L h_{\mu\nu} -2\Lambda h_{\mu\nu}=0.
\eeq{16}
Gauge invariance thus removes four degrees of freedom. We can still perform 
gauge transformations that preserve the gauge fixing 
Eq.~(\ref{15}):
\beq
0=\nabla^{\mu} \nabla_{(\mu} V_{\nu)} - {1\over 2} \nabla_\nu \nabla^\mu
V_\mu=
{1\over 2} \nabla^2 V_\nu +{1\over 2} \nabla^\mu \nabla_\nu V_\mu
- {1\over 2} \nabla_\nu \nabla^\mu V_\mu= {1\over 2}(\nabla^2 +\Lambda)V_\nu.
\eeq{17}
This residual gauge invariance removes four on-shell degrees of freedom, 
since when $V_\mu$ obeys Eq.~(\ref{17}) $\nabla_{(\mu} V_{\nu)}$ solves the
equation of motion~(\ref{16})
\beq
\Delta^{(2)}_L\nabla_{(\mu} V_{\nu)} -2\Lambda \nabla_{(\mu} V_{\nu)}=
\nabla_{(\mu}(\Delta^{(1)}_L -2\Lambda)V_{\nu)}=0.
\eeq{18}
In total, gauge invariance removes $4+4=8$ degrees of freedom, leaving two
physical propagating degrees of freedom.
\section*{One-Particle Amplitude}
Let us proceed now to compute the field produced by a covariantly conserved 
source ($\nabla^\mu T_{\mu\nu}=0$).

First, decompose $h_{\mu\nu}$ as follows
\beq
h_{\mu\nu}=h_{\mu\nu}^{TT} + \nabla_{(\mu} V_{\nu)} +
\nabla_\mu \nabla_\nu \phi +g_{\mu\nu}\psi,\qquad \nabla_\mu V^\mu=0, \qquad
\nabla^\mu h_{\mu\nu} ^{TT}= h^{TT}=0.
\eeq{19}
Using Eq.~(\ref{8}) we find
\beq
\nabla^\mu \nabla^\mu h_{\mu\nu} = \nabla^4 \phi + \Lambda \nabla^2 \phi +
\nabla^2 \psi, \qquad
h =\nabla^2 \phi + 4\psi.
\eeq{20}
Using Eq.~(\ref{20}) to express $\psi$ in terms of 
$\nabla^\mu\nabla^\mu h_{\mu\nu}$ and $h$ , thanks to Eqs.~(\ref{11},\ref{12})
we find
\beq
(3\nabla^2 + 4\Lambda)\psi = - \nabla^\mu \nabla^\mu h_{\mu\nu} + (\nabla^2
+\Lambda)h=\Lambda h = - {4\Lambda \over 3M^2 -2\Lambda} T;
\eeq{21}
\beq
\psi= {4\over 6-9M^2/\Lambda}(\nabla^2 + 4\Lambda/3)^{-1}T.
\eeq{22}
The one-particle exchange amplitude between two covariantly conserved sources,

$T_{\mu\nu}$ and $T'_{\mu\nu}$ can now be written very simply as
\beq
A={1\over 2} \int d^4 x \sqrt{-g} T'_{\mu\nu}(x) h^{\mu\nu}(x)\equiv
{1\over 2} T'_{\mu\nu} h^{TT\, \mu\nu} + {1\over 2} T' \psi.
\eeq{23}
The transverse traceless part of Eq.~(\ref{2}) is
\beq
(\Delta^{(2)}_L + M^2 -2\Lambda) h_{\mu\nu}^{TT} = 4T^{TT}_{\mu\nu}, 
\eeq{24}
\beq
T^{TT}_{\mu\nu}= T_{\mu\nu} -{1\over 3} g_{\mu\nu} T +{1\over 3}(
\nabla_\mu \nabla_\nu + g_{\mu\nu} \Lambda/3) (\nabla^2 + 4\Lambda/3)^{-1}T.
\eeq{25}
Eqs.~(\ref{22},\ref{24},\ref{25}) are all we need to 
compute the one-particle exchange 
amplitude. Using again Eqs.~(\ref{4}-\ref{8}) we obtain
\bea
A &=& 2T'_{\mu\nu} (\Delta^{(2)}_L + M^2 -2\Lambda)^{-1} T^{\mu\nu} 
-{2\over 3} T' (-\nabla^2 + M^2 -2\Lambda)^{-1}T + \nonumber \\
&& +{2\Lambda\over 9} T' ( -\nabla^2 +M^2 -2\Lambda)^{-1} 
(\nabla^2 +4\Lambda/3)^{-1}T + 
{2\over 6 -9M^2/\Lambda} T'(\nabla^2 +4\Lambda/3)^{-1}T .\nonumber \\ &&
\eea{26}
This amplitude seems to have an unphysical pole at $\nabla^2=-4\Lambda/3$, 
but it does not. Indeed, the residue at $\nabla^2=-4\Lambda/3$ is
\beq
\left({2\Lambda \over 9}\right) {1\over 4\Lambda/3 -2\Lambda + M^2} + 
{2\over 6-9M^2/\Lambda}=0\;\;!
\eeq{27}
At the physical pole, $\nabla^2= M^2-2\Lambda$, instead, the residue is
\beq
-{2\over 3} + \left({2\Lambda \over 9}\right) {1\over M^2-2\Lambda/3} =
{2\Lambda -2M^2\over 3M^2 -2\Lambda}.
\eeq{28}
\section*{$M^2\rightarrow 0$, $\Lambda \rightarrow 0$}
Finally, we want to discuss the massless limit as well as the 
$\Lambda\rightarrow 0$ limit. At $M^2=0$, $\Lambda\neq 0$,
the one-particle amplitude is most easily computed in the gauge
$\nabla^\mu h_{\mu\nu}=\nabla_\nu h$. With this gauge choice $h$ and $\psi$
are determined by Eqs.~(\ref{12},\ref{22}) with $M^2=0$. This result is 
obvious since the amplitude in Eq.~(\ref{26}) is nonsingular in the limit
$M^2\rightarrow 0$. More generally, the amplitude is smooth in the limit
$M^2/\Lambda \rightarrow 0$. This is the limit invoked in  ref.~\cite{kr}.

Notice that Eq.~(\ref{26}) reproduces the known vDVZ amplitude in the limit 
$M^2=constant$, $\Lambda \rightarrow 0$. 
Let us examine the massive limit first. In that case, the residue at the 
physical pole has a smooth limit for $\Lambda \rightarrow 0$
\beq
\lim_{\Lambda \rightarrow 0}\left({2\Lambda -2M^2\over 3M^2 -2\Lambda}\right)=
-{2\over 3}, \qquad M^2\neq 0
\eeq{29}
i.e.
\beq
\lim_{\Lambda\rightarrow 0} A = 
2T'_{\mu\nu}(-\Box^2 + M^2)^{-1} T^{\mu\nu} -{2\over 3} 
T'(-\Box^2 + M^2)^{-1}T, \qquad M^2\neq 0.
\eeq{30}
This is the vDVZ amplitude for a massive spin-2 in flat space~\cite{vvdz}.

At $M^2=0$, the residue is always $-1$ so that
\beq
\lim_{\Lambda\rightarrow 0} A = 
-2T'_{\mu\nu}\Box^{-1} T^{\mu\nu} + 
T'\Box^{-1}T, \qquad M^2= 0.
\eeq{31}
This is the one-particle exchange amplitude for a massless graviton in pure 
Einstein's theory~\cite{vvdz}.    

To conclude, we found that the vDVZ discontinuity is an accident of Minkowski 
space, and that it is not present in AdS space. At $\Lambda <0$, the
$M^2/\Lambda \rightarrow 0$ limit is smooth. In particular, whenever 
$M^2/\Lambda \ll 1$, there is little difference between the predictions of 
Einstein's gravity and those of massive gravity.
Experimentally, whenever $M\ll 1/R$, with $R$ the Hubble radius, a massive 
spin-2 is indistinguishable from a massless one. We want to stress again 
that this is {\em not} the case in flat space.

As we mentioned before, 
the absence of a vDVZ discontinuity in {\it de Sitter} space 
was noticed, with a different method, in refs.~\cite{h}. This is also 
apparent in our formalism, since the $M^2/\Lambda \rightarrow 0$ limit is
smooth for either signs of the cosmological constant. In de Sitter space, 
though, this limit is meaningless, since spin-2
representations of the de Sitter group are non-unitary in the range 
$0< M^2 < 2\Lambda/3$~\cite{h}. This pathology manifests itself also in our 
formalism, in the fact that the amplitude Eq.~(\ref{26}) is singular at 
$M^2=2\Lambda/3$.
 
It would be intriguing to conjecture that the peculiar vDVZ 
discontinuity in flat space  may be
somewhat related to the cosmological constant problem, but we
won't. Someone else already said it best: ``hypotheses non fingo.''
   
\vskip .2in
\noindent
{\bf Addendum}
After completion of this paper but before submission to the archives,
ref.~\cite{k} appeared, reaching the same result as this paper.
\vskip .2in
\noindent
{\bf Acknowledgments}\vskip .1in
\noindent
We thank C. Fronsdal, G. Dvali, P. van Nieuwenhuizen and in particular 
L. Randall for useful discussion and comments.
This work is supported in part by NSF grants PHY-9722083 and PHY-0070787.


\begin{thebibliography}{99999}
\bibitem{vvdz} H van Dam and M. Veltman, Nucl. Phys. B22 (1970) 397; V.I. 
Zakharov, JETP Lett. 12 (1970) 312.
\bibitem{bd} D.G. Boulware and S. Deser, Phys. Rev. D6 (1972) 3368.
\bibitem{dgp} G. Dvali, G. Gabadadze and M. Porrati, Phys. Lett. B484 (2000)
112, hep-th/0002190; Phys. Lett. B484 (2000)
129, hep-th/0003054.
\bibitem{grs} R. Gregory, V.A. Rubakov and S.M. Sibiryakov, Phys. Rev. Lett.
84 (2000) 5928, hep-th/0002072.
\bibitem{c} C Csaki, J. Erlich and T.J. Hollowood, Phys. Lett. B481 (2000)
107,
hep-th/0003020.
\bibitem{prz} L. Pilo, R. Rattazzi and A. Zaffaroni, JHEP 0007 (2000) 056, 
hep-th/0004028. 
\bibitem{kro} I. Kogan and G. Ross, Phys. Lett. B485 (2000) 255, hep-th/0003074.
\bibitem{kr} A. Karch and L. Randall, to appear.
\bibitem{rs} L. Randall and R. Sundrum, Phys. Rev. Lett. 83 (1999) 3370, 
hep-ph/9905221; Phys. Rev. Lett. 83 (1999) 4690, hep-th/9906064. 
\bibitem{gkr} S.B. Giddings, E. Katz and L. Randall, JHEP 0003 (2000) 023,
hep-th/0002091. 
\bibitem{h} A. Higuchi, Nucl. Phys. B282 (1987) 397; ibid. B325 (1989) 745.
\bibitem{pf} M. Fierz, Helv. Phys. Acta 12 (1939) 3; 
M. Fierz and W. Pauli, Proc. Roy. Soc. 173 (1939) 211. 
\bibitem{ad} L.F. Abbott and S. Deser, Nucl. Phys. B195 (1982) 76. 
\bibitem{l} A. Lichnerowicz, ``Propagateurs et commutateurs en relativit\'e
g\'enerale,'' Institute des Hautes Etudes Scientifiques, Publications 
Math\'ematiques, 10 (1961).
\bibitem{k} I.I. Kogan, S. Mouslopoulos and A. Papazoglou, hep-th/0011138.
\end{thebibliography}
\end{document}